\documentclass[english,12pt,a4paper]{iopart}

\usepackage[english]{babel}
\usepackage{amssymb}
\usepackage{amscd}
\usepackage{graphicx}
\usepackage{epsfig}

\newcommand{\beq}{\begin{equation}}
\newcommand{\enq}{\end{equation}}

\newcommand{\TT}{\mathbb{X}}

\newcommand{\sech}{\mbox{sech}}

\def\Journal#1#2#3#4#5{#1 #5 {\it #2} {\bf #3} #4}
\def\Book#1#2#3#4#5{#1 #5 {\it #2} (#4: #3)}

\def\eq{\equiv}

\def\g{\gamma}

\def\s{\sigma}

\begin{document}

\title[On the effect of random inhomogeneities in Kerr-media modelled by NLS equation]{On the effect of  random   inhomogeneities in  Kerr-media modelled by non-linear Schr\"{o}dinger equation}
\author{Javier Villarroel$^{1}$ 
and Miquel Montero$^{2}$}
\address{$^{1}$ Facultad de  Ciencias, Universidad  de Salamanca,     Plaza Merced s/n, E-37008 Salamanca, Spain}  
\address{$^{2}$ Departament de F\'{\i}sica Fonamental, Universitat de Barcelona, Diagonal 647, E-08028 Barcelona, Spain}
\eads{\mailto{javier@usal.es}, \mailto{miquel.montero@ub.edu}}

 \begin{abstract}

   We consider   propagation of  optical
    pulses   under
the interplay of dispersion and Kerr non-linearity  in
optical  fibres  with impurities    distributed at  random    uniformly on the fibre.      By using a model based on the
  non-linear Schr\"{o}dinger equation we clarify how
such   inhomogeneities   affect different
aspects such as  the number of solitons present and  the intensity  of the signal.   We also obtain the mean distance
 for the signal to dissipate to a given  level.

\end{abstract}
\pacs{42.81.Dp, 42.65.Sf, 02.50.Ey}
\maketitle

 \section{Introduction} In this article  we consider the   evolution of  a complex electric field $u(x,t)$ in a  non-linear Kerr media
      which has constant dispersion and  losses
  and,
    in addition, impurities at certain points $x_n$, $x_n <x_{n+1}$, which {\it
occur randomly } on the fibre.    We suppose that  these   loss elements    cause the ``input"
      signal $u(x_n^{-},t)$
      to
abruptly decrease       to an   ``output" value
     $u(x_n^{+},t)= e^{-\g_{n}} u(x_n^-,t)$, where  $e^{-\g_{n}}<1$  measures  the   dimming ratio and $ u(x_n^-,t)$, say, denotes
      the limit value from the left.  Assuming the validity  of the
     self-focusing non-linear  Schr\"{o}dinger (NLS) equation as a model of ideal transmission \cite{H} we find that
     the above situation must be described  by a perturbed NLS equation 
       which written in
dimensionless units reads
\begin{equation}  iu_x+ u_{tt}+ 2|u|^2u=    i
    \left[ -\Gamma u+\sum_{n}\big(e^{-\g_{n}}-1\big) \delta(x- x_{n}) u(x_n^-,t )\right],
\label{Eq1}
\end{equation}
  where the Dirac-delta terms   account precisely for the amplitude decrease at impurities; further
   $\Gamma\ge 0$ is the normalized loss coefficient.
For the sake of avoiding extra mathematical difficulties we do not consider a compensated
loss mechanism; this will be the subject of a future publication. We also remark that with
minor changes our results may be applicable to other physically interesting systems such as
Bose-Einstein condensates.

It appears that while the effect of {\it continuous} random   noise |or white noise| on NLS solitons has been well studied in the literature
  (see \cite{Elgin,Gordon,Menyuk,Wadati,Villarroel2})  far less is
 known as regards the effects of sudden, discrete random perturbations.    We intend to  clarify  how these  inhomogeneities |which may be  relevant for
long-distance fibre-optic communication systems|
     affect   the evolution  of the pulse.  We remark that perturbations involving delta masses also appear  related
     to  erbium-doped amplifiers  and  dispersion management,  see \cite{HK0,ABCJS,ABO}.   In such a context, the positions   of the amplifiers $x_n$   are   {\it
 deterministic}  and      periodically disposed,    $x_n\eq n x_1$,   while the strengths are    constant and  negative,   $\g_{n} = - \Gamma x_1 $.     Kodama  and Hasegawa  \cite{KH3}
generalize the latter  ideas to a random context but, unlike us,
maintain the amplifier interpretation and   consider  the
distribution of the ``intensity'' of the signal only in the limit when both
$\Delta_n$ (here  $\Delta_n\eq x_n - x_{n-1}>0$  is the distance between impurities) and $\g_{n}$ 
tend to zero. Thus while these ideas
have some bearing with our work both the physical interpretation and the mathematical model are quite different.

We will start our analysis of equation~(\ref{Eq1}) by considering that there are no deterministic losses, $\Gamma=0$, since this case is simpler from a mathematical viewpoint: We show that   upon  performing a change of
dependent variable the resulting formula can be
piecewise related to the unperturbed NLS   equation.
Let us recall here that the classical NLS  equation  
\begin{equation}i\Theta_x+ \Theta_{tt}+
2|\Theta|^2\Theta= 0, \
\Theta(0,t)=\varphi(t),
\label{NLS}
\end{equation} 
was first derived
by Zakharov \cite{Zakharov2}  as an equation of  slowly varying wave packets of small
amplitude.  He     showed   that  despite its non-linear character  the
corresponding  initial value problem (IVP)  can be reduced to a linear problem (the Zakharov-Shabat spectral problem)  by the so called
inverse scattering transform (IST) |see
\cite{Ablowitz-Segur,Ablowitz-Clarkson} for general background on NLS equation
and the IST method.     Its  interest
    has been  further underlined by  the realization     that
   it  also models
the evolution of the complex amplitude of an optical
    pulse   in a non-linear
  fibre \cite{H}. Applications of NLS equation to optical communications  and  photonics are nowadays standard \cite{H,HK0,Hasegawa,HK2}.
We devote section~\ref{S2} to the study of   the non-linear dynamics
of the classical solitary waves within this regime, and we will show  how    impurities result in the appearance of radiation and general broadening of the signal. In particular, we find that
 solitons may be destroyed by the action of just   one impurity.

  When $\Gamma\ne 0$  equation~(\ref{Eq1})    is no longer solvable in analytic way by IST; however we find |see section \ref{S3}| that the evolution of
          intensity, momentum and position   of the pulse can be described precisely  and  that, under certain natural assumptions,
           their average values       {\it     decrease
    exponentially} due to the  ``impurities'': concretely,  we  suppose that    positions and strengths    of impurities
    are statistically independent  between themselves; we also suppose that in any interval $[0,x]$ impurities are uniformly distributed
    (provided its   number is given).    Nevertheless  the frequency and position of the   pulse are not affected.

In section \ref{S4} we   study  the mean distance for the signal's intensity     to attenuate to a given level due to the impurities. In applications,
this level could be a recommended threshold value for reliability,
say. To this end we
  formulate a linear  integral equation that  this distance satisfies  and,  by means of a Laplace
 transform, solve it.  Results are discussed.

\section{Method of solution and the loss-less case}
\label{S2}

\begin{figure}[hbtp]
{\hfil\includegraphics[width=0.85\columnwidth,keepaspectratio=true]{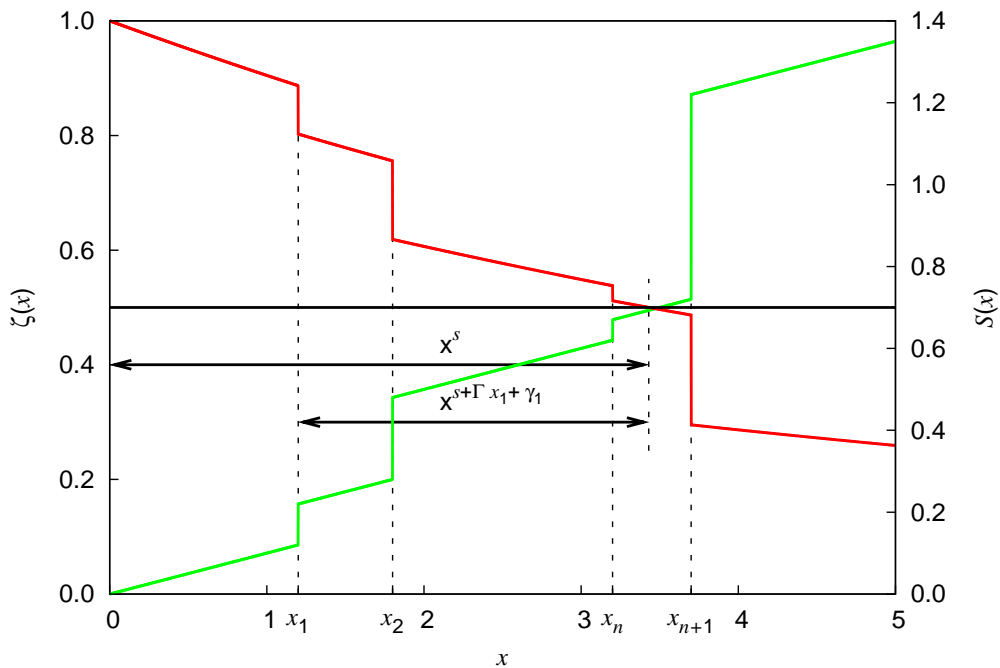}}
\caption{A sample path of   $\zeta (x)$ (in red) and  $ S(x)$
(green line) versus distance (in Km) showing the
 distance $\mathbf {x}$ for the energy to dissipate to half its initial
 value.   We take a fibre with   mean   impurities  distance  $ <\Delta_n>=\lambda^{-1} =1$
km, a loss rate $0.02$dB/Km and dispersion distance $50$ km,
i.e., $\Gamma=0.1$ |which   accounts for the  seemingly linear
behaviour      between
 jumps.} \label{Fig1}
\end{figure}

  Here we  solve  (\ref{Eq1})  given  arbitrary sequences $x_n$ and $\g_n$ with $0<x_n<x_{n+1} $ and $  \g_n>0 $.
  We perform the change of variable    $u(x,t)=
    \zeta(x)\upsilon(x,t)$ where we require that $\zeta(x)$ depends
    {\it only on  space and has
 jump discontinuities} at $x_{n}$ and that
 $\upsilon(x,t)$  \it be  continuous\rm .
    By substitution we find that these functions must solve  the equations
    \begin{eqnarray} i\upsilon_x+\upsilon_{tt}+  \zeta^2
|\upsilon|^2\upsilon= 0, \label{Eq2}\\
\frac{d \zeta(x)}{dx}   +
 \Gamma\zeta  (x)+\sum_{n} \big(1-e^{-\g_{n}}\big) \delta(x- x_{n})\zeta
(x_n^-)   =0.    \label{Eq3}
\label{Eq3}
\end{eqnarray}
 It follows  that $\zeta(x)$ is
continuous on the intervals $(x_n,x_{n+1})$ wherein it solves equation
(\ref{Eq3}) with no delta terms; further, it has jump
discontinuities
 at the random points
$x=x_n$  at which $\zeta(x_n^+)=  e^{-\g_{n}}\zeta(x_n^-)   $.
Hence   if  $N(x)$  is the number of
 defects   on $[0,x]$ we have that
\begin{equation}
\zeta(x)=   e^{- S(x)}, \mbox { where
   }\  S(x)\eq s+ \Gamma   x + \sum_{j=1}^{N(x)} \g_j.
\label{Eq5}
\end{equation}
 Alternatively,  $S(x)=s+ \Gamma   x + \sum_{j=1}^{ n} \g_j,\  $ if $x_n\le x<x_{n+1}$.
(By contrast, for erbium-doped fibre amplifiers $ S(x)= -    \Gamma x -n\Gamma x_1    $ if  $nx_1\le x<
 (n+1)x_1$, see \cite{ABCJS}.)
Thus $ S(x)$ is a piece-wise linear function with initial value $s$ and jumps at the random points $x=x_n$,
i.e., a pure random point process with drift, well known in the
 physics literature.  For convenience we take    $s=0$ hereafter, and until section \ref{S4}. There we will need to consider a more general situation
 where   the starting value $S(0)$ is free.
In figure~\ref{Fig1} we plot a sample of both   $ S(x)$   and 
$\zeta(x)$ for a particular choice of the parameter set under this assumption.

We shall now focus  our  attention  in  equation (\ref{Eq2}).
We first consider the simpler  case when  the loss vanishes:  $\Gamma=0$. It turns out that, even
  though the resulting equation has random discontinuous   coefficients,   it
can be   piecewise reduced  to an integrable equation     whereupon we  show how to obtain the evolution of an initial pulse
(see  \cite{Lund} for related considerations).
 The reasoning in the rest of this section is essentially independent of the sequences $   \Delta_n\equiv x_n-x_{n-1} $ and $     \g_n  $.
 Nevertheless we
shall suppose that          both
are sequences of {\it positive, independent, equally distributed
} random variables and    that       $ \Delta_n  $  and
   $   \g_m $ are also  independent for all $n$, $m$.       Note that all these assumptions
   are   physically well founded  as  they  imply, say,  that the knowledge of the position of  a given
   impurity does not provide any information on the location of the remaining ones.
  The further assumption
that $ \Delta_n $
   is \it exponentially distributed: \rm   $\Pr\Big(\Delta_n  \ge
x\Big)= e^{-\lambda x}$ where $\lambda \eq <\Delta_n>^{-1} $   is
a certain
   parameter,  is  natural from physical
   principles. It  has several fruitful consequences as then there follows that the number   $N(x)$ of impurities
   that occur on $[0,x]$   has
Poisson
   distribution with parameter $\lambda x$
     and that they are uniformly
   distributed on the interval. It further implies the memory-less property:   the distribution of impurities on $(x,x+\Delta x]$
    remains unaffected  given that
     none was    observed on $[0,x]$.  By contrast, we consider here a general
     probability density function (PDF)    $h(y)$ of $\g_n$: $\Pr
\big(y<  \g_n\leq y+dy\big) =h(y)dy$.

For the sake of being specific let us consider the case when the
initial data is just  a   solitary wave  pulse:
$ \upsilon(0,t)=2\eta\, \sech\left(2\eta   t  \right) e^{ 2i  \xi t }\equiv\varphi^{(0)}(t)$
where   the  real parameters $\eta$ and $\xi$  give,  up to  a constant,   the wave's amplitude and      the carrier   velocity.~\footnote[1]{We adopt the
    convention  and terminology of standard   NLS theory wherein $t$ is space and $x$ a temporal variable,
    a situation opposite to that that occurs in Optics.} Note that  up to the first impurity    $ \upsilon^{(0)}(x,t)\eq
\upsilon(x,t),  0\le x \le x_1$, solves  the 
IVP
\beq i\upsilon_x^{(0)}+\upsilon^{(0)}_{tt}+2
|\upsilon^{(0)}|^2\upsilon^{(0)}= 0, \
 \upsilon^{(0)}(0,t)=\varphi^{(0)}(t).
 \label{UNLS}
 \enq
 This is  the standard  IVP for  NLS equation and hence
the solution for $0\le x \le x_1$ 
is the classical soliton
 \beq
 u(x,t)=\upsilon^{(0)}(x,t)=2\eta\,\sech\left(2\eta(t  -4\xi x)\right) e^{ i\left[
2\xi t    +4(\eta^2-\xi^2)x \right]}.
 \label{sol}\enq
As commented, we continue this solution to the interval $x_1 \le x \le x_2$ by requiring $\upsilon(x,t)$ to be continuous at $x=x_1$.
This requirement fixes $\upsilon^{(1)}(x,t)\equiv\upsilon(x,t)$, $x_1\le x \le x_2$, to satisfy the non-linear partial differential equation 
\begin{eqnarray*}    
    i\upsilon_x^{(1)}+\upsilon_{tt}^{(1)}+ 2e^{  -2\g_1}
|\upsilon^{(1)}|^2\upsilon^{(1)}= 0, \mbox{ with} 
\\   \upsilon^{(1)}  (x_1,t)= 2\eta\,\sech\left(2\eta(t  -4\xi x_1)\right) e^{ i\left[
2\xi t    +4(\eta^2-\xi^2)x_1 \right]}. 
\end{eqnarray*}
Remarkably {\it this equation can be reduced again to NLS}: by using both temporal and translational invariance of NLS equation one can prove that
\begin{equation*}
e^{\g_1}u(x,t)=\upsilon^{(1)}(x,t)=e^{  4i(\eta^2+\xi^2)x_1 + \g_1} \Theta   (x-x_1,t-4\xi  x_1),
\end{equation*}
where  $\Theta(\cdot,\cdot)$
is the solution to  the NLS equation~(\ref{NLS}), with data at $x=0$ given by $\Theta(0,t)=   e^{  - \g_1}  \varphi^{(0)}(t)$.
Notice that, unlike $\upsilon (x,t)$, $u(x,t)$ is not continuous  at $x=x_1$.

 The determination of the specific form of the solution requires
  solving
a linear spectral  problem.  The procedure is awkward but fortunately       the solution's  main features   may to a large extent be determined
avoiding these complexities.
 We note
   that  if $\g_1\ne 0$   the solution that evolves from   data $ \Theta(0,t)=   e^{  - \g_1}  \varphi^{(0)}(t)  $      is no longer a soliton
    but a complicated pulse
 that may contain     radiation, in addition to the soliton. The former   component
      has a much weaker  rate of decay than the later; concretely, it  decays  as    the corresponding
       solution for the  linearized  Schr\"{o}dinger equation
    (i.e. as $t^{-1/2}$, see
\cite{Ablowitz-Segur}).  Further, if   $\g_1\gtrsim 1.41$ the
arriving soliton  at $x=x_1$ |cf. equation (\ref{sol})| is destroyed by
the action of the fist impurity  after $x_1$.~\footnote{This stems from the fact
that the
 condition
 $\Xi^2I_{0}\left( 2  \Xi \right) <1  $  on the   initial data    guarantees  that no    solitons will   be formed upon evolution \cite{Ablowitz-Segur,VAP}.  Here $I_0(\cdot)$ denotes the modified Bessel
function of zero order and   $\Xi \eq \int_{-\infty}^{\infty} |\Theta(0,t)|dt=\pi e^{  - \g_1}
$.}   Hence
      the  resulting configuration for $x>x_1$ consists {\it solely of radiation}. To be  specific, suppose
  that the  jump  PDF $h(\cdot)$ has exponential distribution with mean $1/\s $.
   Then, after the first impurity the soliton  disappears  with  probability bounded below by    $\Pr  (\g_1\ge y)=e^{-\s y}$, where $y=1.41$.

    Finally, we mention that by using similar ideas one  can extend the solution
to $x>x_n$ by solving (\ref{Eq2}) with data $\upsilon^{(n-1)}(x_n^-,t)$, where as before $\upsilon^{(n)}(x,t)$ denotes the general solution $\upsilon(x,t)$ restricted on $x_{n-1}\le x\le x_n$.
 Translation invariance allows one to reduce this to NLS equation with new data which involves a   contraction  factor  $ e^{  - (\g_1+\dots+\g_n)}$.
  Eventually, this  dimming of the  initial  signal
results in a disappearance of the starting solitons into radiation, an indication  that, as a result of impurities, broadening of the signal takes place.
 We skip  the mathematical  details.

\section{General case with deterministic loss and impurities}
\label{S3}

When $\Gamma>0$    equation~(\ref{Eq2}) can be mapped into  the  so called dispersion-managed NLS equation, which,
   unfortunately,  is  not
 solvable in   analytic way, neither by  using  IST nor by any other method.
   It is then remarkable that  {\it the evolution of the     main    physically observable functionals can be discerned in an exact way}.
  Consider the following quantities

\begin{eqnarray*}
M(x)\eq \int_{-\infty}^{\infty}   |u(x,t)|^2\,dt,\\
P(x)\eq  i\int_{-\infty}^{\infty} \bar{u}(x,t)u_t(x,t)\,dt, \mbox{ and}\\
Q(x)\eq \int_{-\infty}^{\infty}  t |u(x,t)|^2\,dt, 
\end{eqnarray*}
where  $ M (x)$ and $P(x) $ are     the
(accumulated) intensity and momentum of the signal at a position $x$, while  $Q(x)/
M (x)\eq T(x)$  is the pulse   position. The functional   $  P(x)/
M (x)\eq\Omega(x)$   is interpreted as the pulse-centre frequency.   The singular nature of the delta terms    prevent us from
determining  the relevant  evolution by  manipulating   equation~(\ref{Eq1}).  Nevertheless, one can rely again   in  the decomposition
  $u(x,t)=\zeta(t)\upsilon(x,t)$ and use equation   (\ref{Eq2}). Then, proper manipulation  of the latter expressions yields  that
\begin{eqnarray*}
M (x) =  M (0)  e^{- 2S(x)} ,\\   P(x) =  P(0)   e^{- 2S(x)},  \mbox{ and}\\  Q(x)= \left[ Q(0)-2  P(0)x\right] e^{-2S(x)}.   
\end{eqnarray*}
Thus the effect of   the presence of impurities
   results in the addition of a multiplicative
 {\it  random factor}  $ e^{- 2S(x)} $ in both intensity and momentum. Note however   that $ \Omega(x)=\Omega(0)$ and $T(x)=\left[T(0)-2  \Omega(0)x\right] $, and hence that
  inhomogeneities have  no effect whatsoever on   position and frequency, a fact that
 accords with the physical intuition.

It is therefore of   interest to
      evaluate    the mean  amplitude's   decrease.
       We do so by first
assuming  that previously $n$ defects have occurred:
   $N(x)=n$.
   Let $\Bbb   E$ denote   statistical averaging  and $\Bbb   E \Big(     \zeta^2(x) | N(x)=n  \Big)$ be  the mean value
   of    $  \zeta^2(x)$   knowing that exactly  $n$
jumps  have  occurred   on $[0,x]$.   Note that given this
information one has  $S(x) = \Gamma x + \sum_{j=1}^n
 \g_j $: i.e.,  only the uncertainty regarding the value of
the $\g_j$'s remains but not that associated with the number of
summands $N(x)$. In view of the assumed
 statistical independence  we have that the mean factorizes as
\begin{eqnarray*}
\Bbb  E \left(\left. \zeta^2(x) \right| N(x)=n \right)=    \Bbb E \left(  e^{-2 \Gamma   x}
\prod_{j=1}^n e^{-2\g_j} \right)   \\=      e^{-2 \Gamma   x}
\prod_{j=1}^n\Bbb E \Big( e^{-2\g_j} \Big ) \noindent =  e^{-2
\Gamma x}
 Q_2^n,   
 \end{eqnarray*}
 where $Q_r\eq  \Bbb E \left[  \exp\big(-r\g_j\big)\right] = \int_0^{\infty}  e^{-r y} h(y)dy<1$ is the Laplace Transform of the jump-size PDF.
 The mean   intensity    is obtained  by further averaging     with respect to the number of
   impurities:
 \begin{eqnarray}  \Bbb  E \big[M(x)\big]= M_0\Bbb  E    \big[ \zeta^2(x)\big] \nonumber \\=M_0   \sum_{n=0 }^\infty
{\big(\lambda x\big)^n e^{-\lambda  x}\over
 n!} \Bbb E  \left(\left.\zeta^2(x)\right| N(x)=n \right)   = M_0e^{  -\left[2 \Gamma
+\lambda(1-Q_2)\right]  x  },
\label{Eq6}
\end{eqnarray}
where  $M(0)\equiv M_0$ and we used that if $\Delta_j$ has exponential distribution, i.e., if $\Pr(\Delta_j\ge x)=e^{-\lambda x}$
for some $\lambda>0$, then $N(x)$, the number of defects on $[0,x]$,
is Poisson distributed: 
$\Pr\left(N(x)=n \right)={\left(\lambda x\right)^n}e^{-\lambda x}/n!$. Hence we obtain that the existence of defects implies an additional  exponential decrease in the field's intensity
and momentum at a rate $2\lambda(1-Q_2)$, an effect which might result in the     degradation  of the bit patterns.

\section{Mean half life}
\label{S4}

  A natural related  problem of interest  is   determining  the    distance $\mathbf {x}$  at which
   $M(x)$
  dissipates from a starting value $M_0$  to a  given level  $ M_1  $, i.e., such that $ M(\mathbf {x}) = M_1 $.
  For convenience we set $ M_1\eq M_0e^{-2b} $ and hence require
    $S(\mathbf {x})=b
  $. This distance   could be considered as a threshold value below which the signal is no longer
  reliable (it gives the mean half life of the signal if
  $M_0=2M_1$).
  In the   deterministic  case ($\lambda=0$) this distance follows   inverting  $M_1= M_0\exp{  \big(- 2 \Gamma
   \mathbf {x}}\big)$ as  $  \mathbf {x}= \frac{1  }{2\Gamma
 }\log\frac{M_0}{M_1}$.
When inhomogeneities are
  present    $\mathbf {x}$
  is a  random variable whose
  mean {\it is not obtained}
   by inverting equation (\ref{Eq6}) |as it might have been naively thought.
   Instead, we reason as follows:
 call $\mathbf {x}^s$, see figure~\ref{Fig1}, the
(random) distance that takes for the generalized process
$S(x)$ with initial value $S(0)=s$ |cf. equation (\ref{Eq5})| to  go beyond the level
  $b$. It turns out  that $\TT_{}
(s)\eq \Bbb E (\mathbf {x}^s)  $ satisfies the {\it  linear
integral equation}  
\begin{equation}
\TT(s)=   \frac{1-e^{-\lambda
\varrho}}{\lambda}+
 \frac {\lambda} {\Gamma}\int _{0}^{b-s} dl  e^{{\lambda \over
\Gamma}( s+l-b)} \int_{0}^{l} dy\TT( y+b-l) h(y),
\label{Eq10}
\end{equation}
 where $\varrho\eq{b-s\over \Gamma }$ and we  recall  that  $h(x)$  is the density of $\g_n$.~\footnote[7]{We sketch
the
 derivation of this integral equation (see \cite{mmp05} for a similar derivation in  a financial context).
  With
   $S(0)=s$ there are three possibilities for the future evolution:  If
 the first jump satisfies   $  x_1  >  \varrho $  then $S $ reaches the level $b$ at   $x= \varrho $. If
this is not the case and if  the jump at $  x_1 $  satisfies $
s+\Gamma   x_1+\gamma _1 \ge b $ then   the process  goes past $b$
at $x=x_1$. Otherwise  the  process still remains within $[0,b)$
 at $x=x_1$  and starts afresh with an initial value $S(x_1)=s+\Gamma x_1+\g_1<b$ (hence the process will exit   $[0,b)$
at $x_1+  \mathbf {x}^{s+\Gamma x_1+\g_1}$). 
Upon appropriate rearrangement this reasoning leads to
   $$  \mathbf {x}^s = \varrho    \theta\big(  x_1- \varrho
   \big)+  x_1\theta\big(   \varrho  -x_1\big) +\mathbf {x}^{s+\Gamma x_1+\g_1} \theta\big(b-s-\Gamma
x_1- \g_1 \big) \theta\big(   \varrho  -x_1\big).  
$$
Averaging this relationship yields with further manipulations equation (\ref{Eq10}).}

\begin{figure}[hbtp]
{\hfil\includegraphics[width=0.85\columnwidth,keepaspectratio=true]{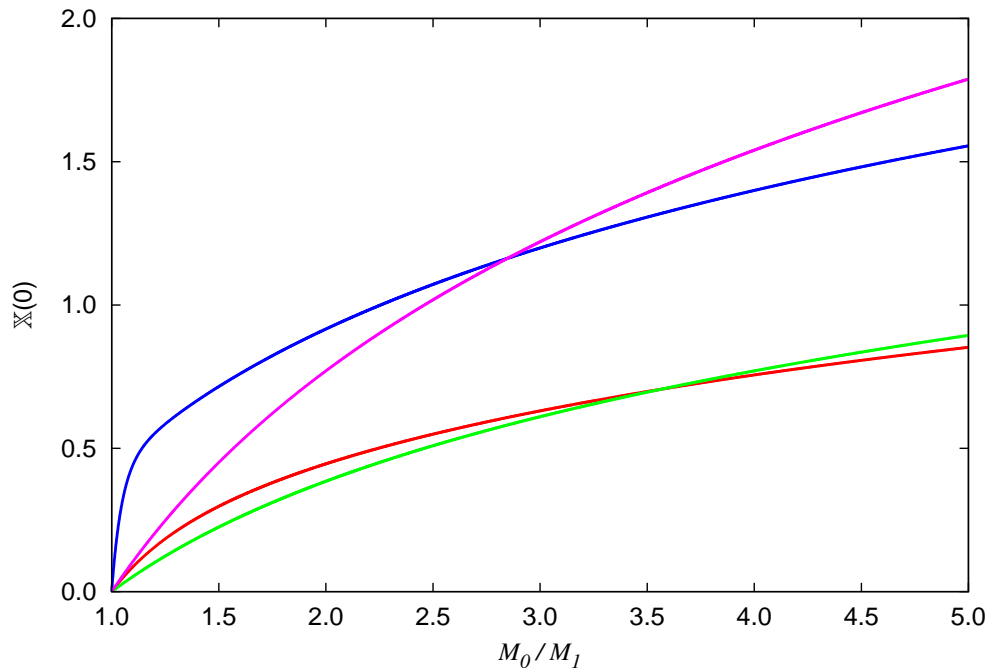}}
\caption{ Mean distance  in terms of $M_0/M_1$ for
$\lambda=2.0$, $\s=3.0$ while  $\Gamma=0.5$
 (red line)  and  $\Gamma=0.05$ (blue one) as follows from~(\ref{Good_x}). Note how in the
 latter case  $\TT(0)$ jumps an amount $<\Delta_n>=0.5$ right after the origin. The green and magenta curves
  are the (incorrect) mean distances implied by equation (\ref{Bad_x}) with the
  above parameters.}
\label{Fig2}
\end{figure}

 This equation  can be  solved in a closed form by Laplace transformation.  We   consider  again the case
 corresponding to a jump  PDF also exponential with mean $\s^{-1}\eq <\g_n>$, i.e., $h(x)=\s e^{-\s
x}$ where $\s>0$. If  $
\kappa = \lambda +\s \Gamma$, Laplace transformation yields the
solution to  (\ref{Eq10}) as
  $$\TT(s)= \frac{\s\Gamma\varrho}{\kappa    }+
\frac{ \lambda}{\kappa ^2  }  \Big( 1- e^{- \kappa   \varrho
 }  \Big).
 $$
  The mean  distance   for
 the    amplitude   to decrease to    $M_1$ follows letting  $s=0$ and $b= \frac{1}{2}\log \frac{M_0}{M_1}$ as
    \begin{equation} \Bbb E(\mathbf {x})   \eq \TT_{}
 (0)=    \frac{ 1  }{2 (\Gamma+  \lambda/\s )   }\log\frac{M_0}{M_1}+
\frac{ \lambda}{\kappa ^2  }  \left[ 1- \left(\frac{M_1}{M_0}\right)^{
  \frac{\kappa   }{2\Gamma}}  \right].  \label{Good_x}\end{equation}
  If $\lambda=0$ we recover the   deterministic limit above: $  \mathbf {x}= \frac{1  }{2\Gamma
 }\log\frac{M_0}{M_1}$.  Note how the incorporation of impurities corrects this formula in a significant way, cf. equation~(\ref{Good_x}). Another interesting limit
is that of vanishing deterministic loss rate, $\Gamma=0$.
The mean attenuation distance can only be
accounted to the presence of impurities and reads 
  $\Bbb E (\mathbf {x}) =\frac{1}{ \lambda} +\frac{\s }{2\lambda }\log\frac{M_0}{M_1}
$. The first term is the mean time for the first jump at $x_1$ to happen; the logarithmic correction corresponds to the mean time to go beyond the level $b$ after the first jump.
Actually, this rate  rules the mean
dissipation distance whenever $M_0>>M_1$ and $\lambda/\s>>\Gamma$.
  In  figure \ref{Fig2} we perform a plot of this function. Note how, by contrast, the distance implied inverting  equation~(\ref{Eq6}), namely
  \begin{eqnarray}
 \Bbb  E \big(M(x)\big) = M_0 \exp \left[-2x \left(  \Gamma + \frac{\lambda
}{\s+2} \right)    \right], \mbox{ and therefore} 
\\
\mathbf{x}=   \left[2 \left(  \Gamma + \frac{\lambda
}{\s+2} \right)    \right]^{-1} \log\frac{M_0}{M_1}, \label{Bad_x}
\end{eqnarray}
deviates from the correct result, equation~(\ref{Good_x}), and  fails to  capture the sharp behaviour occurring   for $  M_1 \approx M_0$.
The  error  increases as $\Gamma$ decreases.

\section{Conclusions}   We have analyzed   how  the existence of randomly distributed impurities  affects
  the   evolution of  an optical pulse    in a   non-linear Kerr media
     with constant dispersion and  loss.    We  suppose that the unperturbed situation is described  by   NLS equation. When the
      deterministic loss    vanishes it is shown   by changing the
dependent variable  that  the resulting equation can be
piecewise related to the unperturbed NLS equation. The effect of impurities in the non-linear propagation is pinpointed.  In particular we  address
the issue of how
they affect the initial solitons and the possibility to dissipate them into radiation. In the general, non-solvable $\Gamma\ne 0$ case we show that while impurities do not influence the frequency and position of the  signal they  induce an
    exponential decrease  of the main physical observables  intensity and momentum and hence a     general  degradation.
We also determine the      mean half life or mean  distance for the signal   to
dissipate to a given threshold value. We find that this distance satisfies a certain  integral equation.  Its analysis shows that
 impurities result in an important
 decrease  in the mean dissipation distance.  To overcome these effects the addition of
    amplifiers  is  in order. The introduction of such a device
     and the relevant
   statistical
   implications will be considered in a future publication.

\ack

We appreciate conversations with Prof. Y. Kodama and M. A. Ablowitz, which   have helped enhance  this article.
 The authors acknowledge support from MICINN under contracts No. 
FIS2008-01155-E, FIS2009-09689, and MTM2009-09676; from  Junta de
Castilla-Le\'on, SA034A08; and Generalitat de Catalunya,
2009SGR417.


\section*{References}
\begin{thebibliography}{99}
\bibitem{H}
\Journal{Hasegawa A and Tappet F}{Appl. Phys. Lett.}{23}{142}{1973}

\bibitem{Elgin}
\Journal{Elgin J N}{Phys. Lett. {\rm A}}{181}{54}{1993}

\bibitem{Gordon}
\Journal{Gordon J P and  Haus H A}{Opt. Lett.}{11}{665}{1986}

\bibitem{Menyuk}
\Journal{Menyuk C R}{J. Opt. Soc. Am. {\rm B}}{10}{1585}{1993}


\bibitem{Wadati}
\Journal{Wadati M}{J. Phys. Soc. Jpn.}{52}{2642}{1983}

\bibitem{Villarroel2}
\Journal{Villarroel J}{Stud. Appl. Math.}{112}{87}{2004}

\bibitem{HK0}\Journal{Kodama Y and Hasegawa A}{Opt. Lett.}{7}{339}{1983}

\bibitem{ABCJS} 
\Journal{Ablowitz M J, Biondini G, Chakravarty S, Jenkins R B and Sauer J R}{Opt. Lett.}{21}{1646}{1996}

\bibitem{ABO} 
\Journal{Ablowitz M J, Biondini G and Olson E S}{J. Opt. Soc. Am. {\rm B}}{18}{577}{2001}

\bibitem{KH3} 
\Journal{Kodama Y and Hasegawa A}{Opt. Lett.}{8}{342}{1983}

\bibitem{Zakharov2}
\Journal{Zakharov V E}{J. Exp. Theor. Phys.}{35}{908}{1972} 

\bibitem{Ablowitz-Segur}
\Book{Ablowitz M J and Segur H}{Solitons and the Inverse Scattering Transform}{SIAM}{Philadelphia}{ 1981}

\bibitem{Ablowitz-Clarkson}
\Book{Ablowitz M J and Clarkson P A}{Solitons, Nonlinear Evolution Equations and Inverse Scattering {\rm(}London Mathematical Society Lecture Note Series {\rm 149)}}{Cambridge University Press}{Cambridge}{1991}

\bibitem{Hasegawa} 
\Book{Hasegawa A}{Optical Solitons in Fibers}{Springer--Verlag}{Berlin}{1990}

\bibitem{HK2} 
\Journal{Hasegawa A and Kodama Y}{Phys. Rev. Lett.}{66}{161}{1991}

\bibitem{Lund} 
\Journal{Lundquist P B, Andersen D R and Swartzlander G A}{J. Opt. Soc. Am. {\rm B}}{12}{698}{1995}

\bibitem{VAP}
\Journal{Villarroel J, Ablowitz M J and Prinari B}{Acta Appl. Math.}{87}{245}{2005}

\bibitem{mmp05} 
\Journal{Masoliver J, Montero M and Perell\'o J}{Phys. Rev. {\rm E}}{71}{056130}{2005}

\end{thebibliography}
\end{document}